\documentclass[aps, prl, reprint, superscriptaddress, floatfix]{revtex4-1}

\usepackage{amsmath}  
\usepackage{color}  
\usepackage{graphicx}  
\usepackage{hyperref}  
\usepackage{natbib}  
\usepackage{tikz}  

\pacs{52.40.Kh, 52.65.Rr, 52.30.Ex}

\begin{document}

\date{\today}

\title{Electron Presheaths: The Outsized Influence of Positive Boundaries on Plasmas}

\author{B.T. Yee}
\email{btyee@sandia.gov}
\affiliation{Sandia National Laboratories, Albuquerque, New Mexico 87185, USA}
\author{B. Scheiner}
\affiliation{Department of Physics and Astronomy, University of Iowa, Iowa City,
             Iowa 52242, USA}
\author{S.D. Baalrud}
\affiliation{Department of Physics and Astronomy, University of Iowa, Iowa City,
             Iowa 52242, USA}
\author{E.V. Barnat}
\affiliation{Sandia National Laboratories, Albuquerque, New Mexico 87185, USA}
\author{M.M. Hopkins}
\affiliation{Sandia National Laboratories, Albuquerque, New Mexico 87185, USA}

\begin{abstract}
    Electron sheaths form near the surface of objects biased more positive than the plasma potential, such as a Langmuir probe collecting electron saturation current. Generally, the formation of electron sheaths requires that the electron-collecting area be sufficiently smaller ($\sqrt{2.3m_e/M}$ times) than the ion-collecting area. They are commonly thought to be local phenomena that collect the random thermal electron current, but do not otherwise perturb a plasma. Here, using experiments on an electrode embedded in a wall in a helium discharge, particle-in-cell simulations, and theory it is shown that under low temperature plasma conditions ($T_e \gg T_i$) electron sheaths are far from local. Instead, a long presheath region (27 mm, approximately an electron's mean free path) extends into the plasma where electrons are accelerated via a pressure gradient to a flow speed exceeding the electron thermal speed at the sheath edge. This fast flow is found to excite instabilities, causing strong fluctuations near the sheath edge.

\end{abstract}

\maketitle

\section{Introduction}
    A sheath is the space charge region found near the physical boundaries of most plasmas. The vast majority of sheaths are ion rich because this is what naturally forms as highly mobile electrons charge a surface negatively.Comparatively little is known about electrons sheaths, although they are routinely produced when objects are biased positive with respect to the plasma potential~\cite{Langmuir1929}.

The most common situation is the electron saturation region region of a Langmuir probe sweep, but they arise in many other situations including negative ion sources~\cite{Bacal2006} and electron sources~\cite{Longmier2006}, positive electrodes employed for blob control~\cite{Theiler2012}, particle circulation in dusty plasmas~\cite{Law1998}, and turbulence-induced particle fluxes~\cite{Richards1994}. Electron sheaths are also common in several other situations, including: near highly emitting surfaces~\cite{Campanell2013}, in microdischarges~\cite{Wang2006}, during the high potential phase of the rf cycle in processing discharges~\cite{Mahony1997}, around electrodynamic tethers~\cite{Sanmartin1993}, the lunar photosheath~\cite{Poppe2011}, around wire arrays used for electron temperature control~\cite{Yip2015}, and in scrape off layer control~\cite{Zweben2009}.

In this Letter, results of experiments, particle-in-cell (PIC) simulations and theory are provided showing that electron sheaths form a long electron presheath extending well into the plasma. Furthermore, electrons are accelerated to high velocities in this region, obtaining a distribution that is flow-shifted to a speed exceeding the electron thermal velocity at the sheath edge. This may be considered an electron sheath analog of the Bohm criterion \cite{Bohm1949}. The fast differential streaming between electrons and ions is found to excite streaming instabilities that give rise to strong fluctuations of the boundary layer region. 

Electron presheaths are found to differ in their essential properties from ion presheaths. In particular: (1) the differential potential $(\Delta \phi)$ is much smaller, nominally by a factor of $T_i/T_e$, (2) it is much longer in extent, nominally by a factor of $\sqrt{m_i/m_e}$, (3) electrons are accelerated by a pressure gradient, in contrast to direct electric field acceleration of ions in an ion presheath, and (4) the differential streaming excites instabilities and strong fluctuations. These results promise new insights into the applications mentioned above. Advances in the fundamental physics of electron sheaths may also lead to new applications.

\section{Experiments}
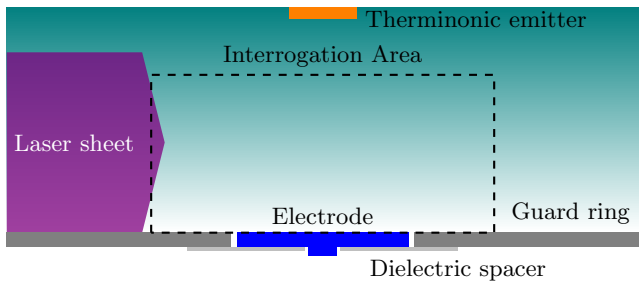
\begin{figure}
    \begin{tikzpicture}[scale=0.6]

\newcommand\RA{7}  
\newcommand\HT{5}  
\newcommand\RE{1.9}  
\newcommand\DL{0.127}  
\newcommand\RI{2*\RE}  
\newcommand\HI{3.5}  
\newcommand\LR{3}  
\newcommand\LH{4}  

\shade [top color={rgb:blue,1;green,1}, bottom color=white]
    (-\RA, 0) rectangle (\RA, \HT);

\fill [orange] (-0.75, \HT-0.25) rectangle (0.75, \HT);
\node [right] at (0.75, \HT-0.25) {Therminonic emitter};

\fill [nearly opaque, blue!50!red]
    (-\RA, 0) --
    (-\RA+\LR, 0) --
    (-\RA+\LR+0.5, \LH/2) --
    (-\RA+\LR, \LH) --
    (-\RA, \LH) --
    (-\RA, 0);
\node at (-\RA+\LR/2, \LH/2) {\textcolor{white}{Laser sheet}};


\fill [blue] (-\RE, 0) rectangle (\RE, -0.3175);
\fill [blue] (-0.3175, -0.25) rectangle (0.3175, -0.5);
\node [above] at (0, 0) {Electrode};

\fill [gray] (-\RA, 0) rectangle (-\RE-\DL, -0.3175);
\fill [gray] (\RA, 0) rectangle (\RE+\DL, -0.3175);
\node [above left] at (\RA, 0) {Guard ring};

\fill [gray!50!white] (0.4, -0.3175) rectangle (3.0, -0.4);
\fill [gray!50!white] (-0.4, -0.3175) rectangle (-3.0, -0.4);
\node [below] at (3, -0.4) {Dielectric spacer};


\draw [dashed, thick] (-2*\RE, 0) rectangle (2*\RE, \HI);
\node [above] at (0, \HI) {Interrogation Area};

\end{tikzpicture}
    \caption{Sketch of the experimental setup used to measure the electron densities above an embedded electrode.}
    \label{fig:experiment_setup}
\end{figure}

Experiments were conducted in a cylindrical vacuum chamber in which an electrode with a diameter of 19 mm was embedded into one face. The walls and faces of the vacuum chamber were maintained at ground potential. The area ratio of the chamber ($1.86\times 10^5$ mm$^2$) to the electrode ($1.14\times 10^3$ mm$^2$) was chosen to ensure the formation of an electron sheath above the auxiliary electrode when biased above the plasma potential \cite{Baalrud2007}. A plasma was generated in 20 mTorr of helium with a barium-impregnated tungsten thermionic emitter (Heatwave Labs model 101117) located 10 cm from the electrode. The emitter was operated at a temperature of approximately $1.1 \times 10^{3}{}^{\circ}$C. A cross section of the experimental setup can be seen in Fig.~\ref{fig:experiment_setup}.

The discharge current was held constant at 300 mA which resulted in a discharge voltage which varied from 49 to 54 V over the course of the measurements. The auxiliary electrode was biased to -50 V, 0 V and 15 V, forming an ion sheath, a weak ion sheath, and electron sheath respectively. An emissive probe was operated at a height of 35 mm and at a radial position of 70 mm and was used to measure the plasma potential via the floating point technique. The plasma potential was 6.2 V, 5.3 V, and 6.6 V for the three aforementioned cases. Two-dimensional maps of the electron densities above the electrode were generated using the laser-collisional induced fluorescence (\textsc{lcif}) diagnostic \cite{Barnat2010}. High energy electrons could potentially cause secondary electron emission, however the uncollided electron density (high energy tail) is estimated to be two orders of magnitude below the measured densities. Given a secondary emission coefficient of $\gamma\approx 0.1$, the density of electrons from secondary emission should be about three orders of magnitudes below the measured densities.

\begin{figure}[t!]
    \includegraphics{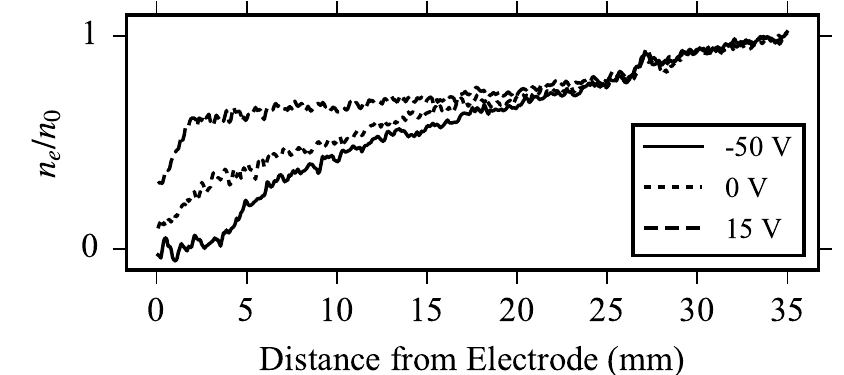}
    \caption{Measured electron density profiles above the electrode along the center axis at -50 V (solid line), 0 V (dotted line), and 15 V (dashed line). The profiles are normalized by their values at 35 mm: $3.5\times 10^{9}$ cm$^{-3}$ at -50 V, $3.3\times 10^{9}$ cm$^{-3}$ at 0 V, and $3.7\times 10^{9}$ cm$^{-3}$ at 15 V. Also included is a gray line representing the constant density gradient found at $z > 30$ mm.}
    \label{fig:on_axis_densities}
\end{figure}

Figure~\ref{fig:on_axis_densities}
shows the axial density profiles from \textsc{lcif} measurements (relative uncertainty estimated to be 7\%~\cite{Weatherford2012b}), normalized by their density at 35 mm, for the ion sheath (-50 V, solid line), a weak ion sheath (0 V, dotted line), and the electron sheath (15 V, dashed line). In all three cases the same linear density gradient is observed far from the electrode ($> 30$ mm).

This gradient is likely the result of the nonuniform generation of plasma in the volume. Attenuation of the primary electrons and their falloff with distance from the small source will lead to a locally peaked density distribution. This non-uniformity will lead to diffusion to the walls of the chamber and an associated ambipolar field. Based on Langmuir probe measurements of the electron temperature and observed density gradients this ambipolar field will be of the order 0.6 V/cm and will tend to inhibit the transport of electrons toward the embedded electrode.

At -50 V and 0 V, the electron depletion region associated with an ion sheath can clearly be seen extending from 0--4 mm and 0--3 mm respectively. The presheath is estimated to be the region at which the electron density gradient deviates from the roughly constant slope ($> 30$ mm). The estimated location of the presheath edge occurs near 12 mm. This can be compared to the calculated ion-neutral mean free path which is expected to determine the presheath length scale~\cite{Riemann1991}. Estimating ion energies of 0.1 eV and using the isotropic scattering cross sections of Phelps~\cite{Phelps2002} ($\sigma = 2.4\times10^{-19}$ m$^2$), we find a mean free path of 6.5 mm which is on the order of the estimated location of the presheath edge.

The electron sheath (15 V) also features a region of electron depletion from about 0--2 mm from the electrode. This is caused by the acceleration of the electron fluid and its consequent rarefaction. The salient feature of the electron sheath case is that its density gradient is essentially the same as the ion sheath's far from the electrode ($> 27$ mm), but closer to the electrode it bears a distinctly different gradient. Assuming that this deviation is caused by collisional effects similar to the ion sheath case, a calculation can be made for electrons based on their momentum transfer cross section. Assuming an energy of 4.0 eV ($\sigma = 6.6\times10^{-20}$ m$^2$ per Phelps~\cite{Phelps2002}), the electron mean free path is found to be 24 mm.

This suggests that the positively biased electrode is affecting the plasma far from the electron sheath region, on the order of an electron mean free path, and suggests the presence of an \emph{electron presheath}. The Debye length at the sheath edge is estimated to be 0.38 mm based on the \textsc{lcif} measurements and Langmuir probe measurements of the electron temperature. Thus this presheath extends approximately 70 Debye lengths from the boundary. In both the case of the electron presheath and ion presheath, the length scales are significantly smaller than the length scale of the system.

\section{Simulations}

A complementary analysis was carried out using Aleph, a PIC-DSMC code developed at Sandia National Laboratories. Aleph is an electrostatic code intended for massively paralllel (over 10k cores) plasma simulations. Fields are solved using finite-element method libraries from the Trilinos project \cite{Heroux2005}. Particles are advanced using the velocity Verlet algorithm as described by Spreiter and Walter \cite{Spreiter1999}. Other recent uses of Aleph include the study of vacuum arc discharges \cite{Timko2012} and the onset of plasma potential locking \cite{Hopkins2016}.

\begin{figure}
    \includegraphics{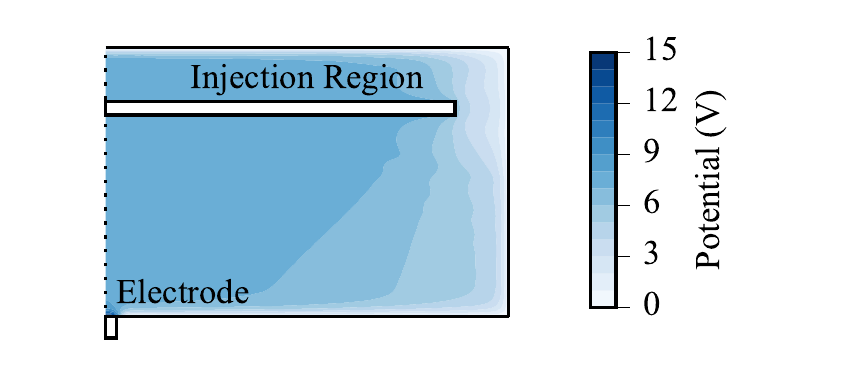}
    \caption{Representation of the simulation geometry with overlaid electric field potential contours from electron sheath simulations. The domain is 75 mm by 50 mm with an axis of symmetry on the left hand side.}
    \label{fig:sketch}
\end{figure}

The simulation was conducted in a rectangular Cartestian domain, 75 mm by 50 mm, with an unstructured mesh formed of triangles. This is the same geometry as that used in~\cite{Baalrud2015} and depicted in Figure~\ref{fig:sketch}.
The left-hand side of the domain specularly reflects all particles and possesses a Neumann boundary condition of $\partial V/\partial x = 0$, thus representing an axis of symmetry. Aside from the electrode shown in Figure~\ref{fig:sketch}, the outer boundaries are held at $V = 0$ and outflux all incoming charged particles. The characteristic size of the triangles used to mesh the domain was 233 $\mu$m which resolved the Debye length of approximately 540 $\mu$m. All simulations used a timestep of 25 ps, which is sufficient to meet CFL requirements for all particles observed in the simulation and resolves the plasma frequency of 0.28 GHz. 

The simulation domain maintains an area ratio of wall to electrode similar to that in the experiment, chosen to assure the formation of an electron sheath \cite{Baalrud2007, Barnat2014}. As the formation of the electron sheath only depends on the area ratio and not the absolute size of the electrode, experiment and simulation should be largely comparable. The simulated electrode was biased to -50 V and 15 V to match the measured ion and electron sheath. Quasineutral plasma ($T_e = 4.0$ eV, $T_i = 1000$ K) was added to the simulation domain at a constant rate in a rectangular area located 37.5 mm from the anode. Simulations were run for 30 $\mu$s at which point they were found to be in equilibrium based on field energy and total particle number. Field and particle properties were averaged over an additional $20$ $\mu$s in order to minimize statistical fluctuation in quantities of interest.

\begin{figure*}
    \includegraphics{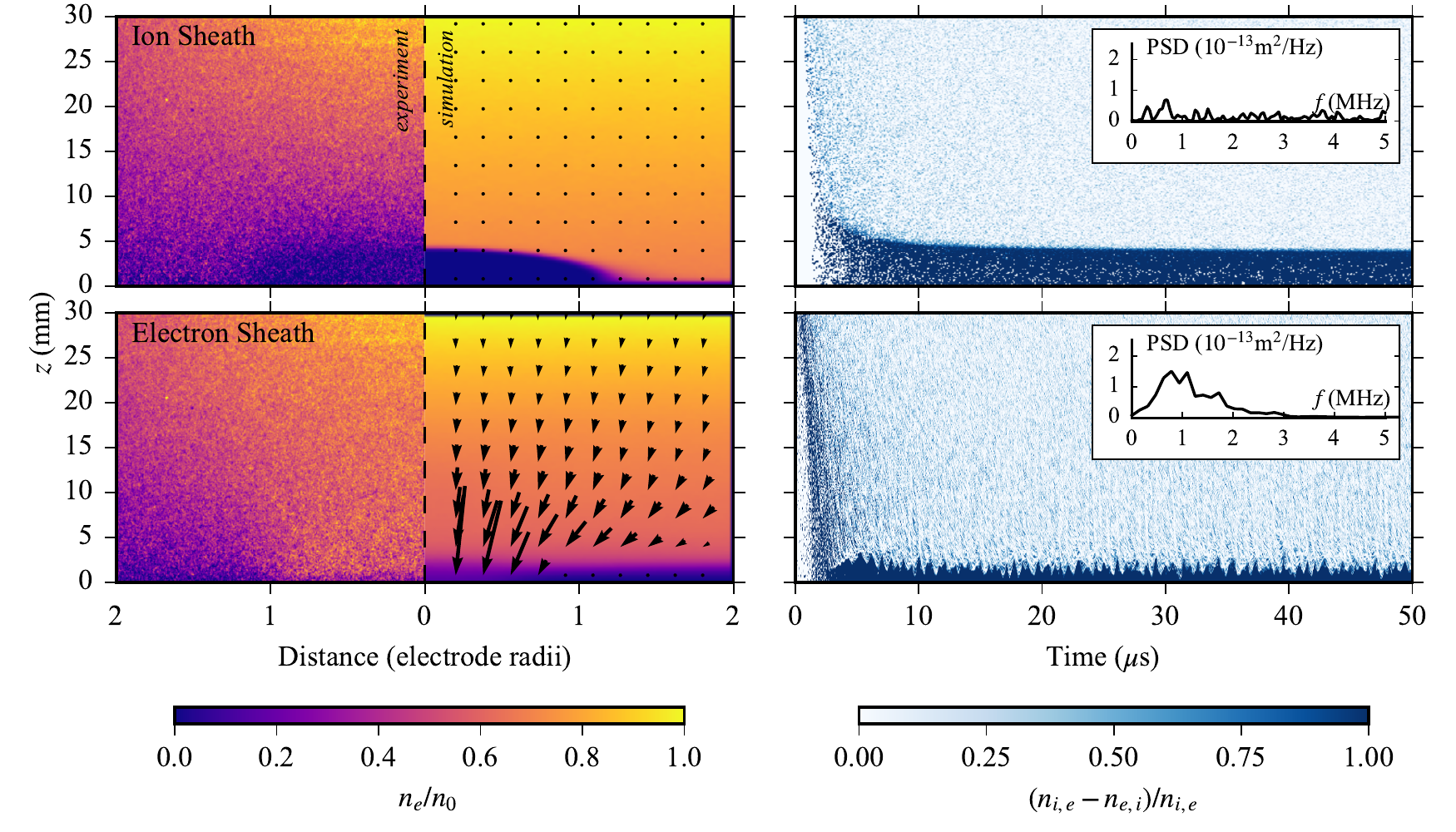}
    \caption{Experimental and simulated electron density maps (left) and simulated on-axis charge densities over time (right). The ion sheath results comprise the upper plots, and the electron sheath results comprise the lower ones. Overlaid atop the simulated electron densities are the electron particle flux, the magnitudes of which range from 0.1--1.2$\times 10^{10}$ m$^{-2}$s$^{-1}$ for the electron sheath, and below $5\times 10^8$ m$^{-2}$s$^{-1}$ for the ion sheath. The power spectral densities of the sheath edge position are inset in the charge density plots.}
    \label{fig:flow_instability}
\end{figure*}

The experimental density measurements are compared to the equivalent simulations in the plots on the left side of Fig.~\ref{fig:flow_instability}.
Overlaid on the simulated electron densities are arrows showing the electron flux vectors scaled to the same value for both cases. The horizontal axes have been normalized by the electrode radii, in order to provide a more suitable comparison. The right side of Fig.~\ref{fig:flow_instability} presents maps of the charge density normalized by the density of the collected species. Inset in the charge density maps are power spectra of the sheaths' positions over time.

In the case of the ion sheath, a large region of electron depletion is visible above the face of the electrode. The size and shape of this region is largely consistent between simulation and experiment, with the small differences likely ascribable to a discrepancy in the electron densities at the sheath edge. An ion sheath is also observed above the grounded wall. Electron current is small throughout the simulated domain and is consistent with the fact that ion sheaths tend to confine electrons. The charge density shows the formation of a stable sheath.


The electron sheath simulation also features a region of electron depletion near the face of the electrode, resulting from acceleration of the electron fluid. Though the width of the sheath in the density maps from simulation appears qualitatively larger in the density maps, the sheath edge seen in the charge densities oscillates about 2 mm consistent with the experimental electron density profile seen in Fig.~\ref{fig:on_axis_densities}. Further differences are likely attributable to a combination of potential causes: lower electron densities in the simulation and the absence of electron-neutral collisions. The former would result in larger Debye lengths and subsequently larger sheaths. The latter would also tend to increase the sheath width as seen in equation (6) of~\cite{Scheiner2015}.

The electron sheath simulation also features a substantial degree of electron current directed toward the electrode from at least as far away as 30 mm. The plasma properties in this region ($n_e=4.5\times10^8$ cm$^{-3}$, $T_e=2.4$ eV) give a Debye length of $0.54$ mm, 60 times smaller than the extent of the directed flow. While it is well known that ion presheaths can extend throughout the entire plasma for large mean free paths~\cite{Riemann1991}, we will show that (for the same collision process) the electron presheath is nominally longer. The extent of the flow is consistent with where the density profile from Fig.~\ref{fig:on_axis_densities} begins to change slope, indicating the presheath. The electron presheath is clearly a significant perturbation to the plasma. Furthermore, as opposed to the ion sheath, the electron sheath edge exhibits significant fluctuations in its position.

The proximity of the grounded wall to the auxiliary electrode leads to a funnel-like structure in the electron flow with a notable convergence. We note that there is the possibility for some overlap of the electron presheath with the ion presheath from the abutting walls. In this case, the radial electric field of the ion presheath will cancel as a result of the symmetry of the system. The axial electric field will tend to accelerate electrons away from, rather than toward the anode, therefore the flows observed in Fig.~\ref{fig:flow_instability} are not the result of adjacent ion sheaths. Finally, we note that the ion presheath length scale is estimated to be 6.5 mm, significantly less than the observed electron presheath size and the device dimensions.

\section{Theory}

Traditional Langmuir probe analysis assumes that a probe in electron saturation collects the random thermal flux of electrons incident on the electron sheath. An implication of this local picture is that the electron velocity distribution function (\textsc{evdf}) at the edge of the electron sheath would be a half-Maxwellian with no flow shift (but with a flow moment)~\cite{Mott-Smith1926, Medicus1961, Hershkowitz2005}. The random flux of electrons flows into the sheath from the quasineutral plasma, and all electrons reaching the sheath are lost to the boundary. However, our results show that the presence of an electron presheath leads to a vastly different picture.  

% ...

\begin{figure}
    \includegraphics{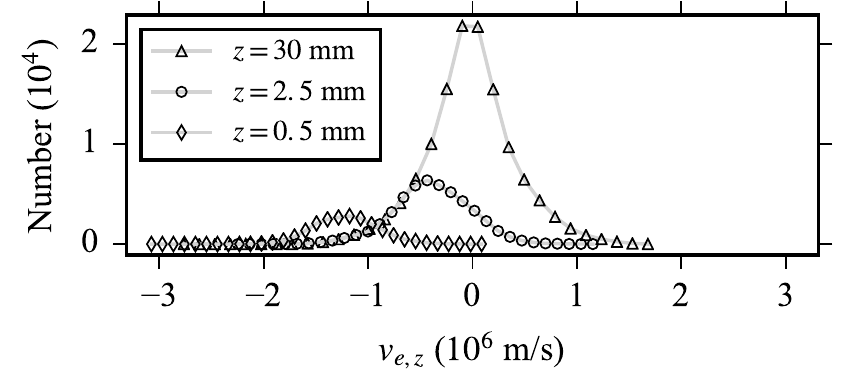}
    \caption{The electron velocity distribution functions normal to the electrode, along $x=0$, at several locations above the electrode in the simulations: the injection region ({\scriptsize $\triangle$}), the sheath edge ($\circ$), and inside the sheath ({\scriptsize $\diamond$}).}
    \label{fig:evdfs}
\end{figure}

In particular, simulations show that the presheath is found to introduce a substantial flow-shift in the electron distribution, approaching the electron thermal speed by the sheath edge; see Fig.~\ref{fig:evdfs}.
Contrary to the conventional picture of a highly-kinetic truncated distribution function, this figure shows that it is in fact well represented by a flowing Maxwellian. That the distribution is well-described by a Maxwellian despite the absence of electron-neutral collisions is notable. Confirmation of this flow shift in experiment remains an open challenge. A significant number of electrons flow out of the electron sheath (negative velocities are electrode-directed) despite the absence of an explicit collision algorithm in the simulations. The sheath is defined as where quasineutrality is sufficiently violated, or $(n_e - n_i)/n_e = \epsilon$ where we have chosen $\epsilon=0.3$ in order to avoid stochastic density variations in the plasma. This approach places the sheath edge at $z = 2.0$ mm. More is said of this choice for $\epsilon$ below, in view of the strong density fluctuations in this region.

While PIC simulations can result in numerical thermalization of nonequilibrium distributions, this is not expected to be a factor in the present results. Estimates based on the work of Montgomery and Nielsen~\cite{Montgomery1970} suggest that the thermalization time for the present system is of the order 2.6 $\mu$s. This exceeds the 1.1 $\mu$s required for the electron fluid to transit from the source region to the electrode. Additionally, recent simulation results of the ion to electron sheath transition~\cite{Scheiner2016} using comparable simulation parameters possess \textsc{evdf}s exhibiting both a flow shift and a loss cone due to the presence of the boundary. Both factors are found to affect the flow speed, with the flow shift being at least as important as the loss cone.

% ...
A fluid analysis is used to interpret aspects of the experiments and simulations and to better understand the origin of the observed flow shift. Consider the fluid momentum balance
\begin{equation}
    u_e\frac{du_e}{dz} = -\frac{e E}{m_e} - \frac{k_B T_e}{m_e n_e} \frac{dn_e}{dz} - u_e (\nu_c + \nu_s)
    \label{eqn:com}
\end{equation}
where $e$ is the elementary charge, $E$ is the electric field, $\nu_c$ is the collision frequency, and $\nu_s$ represents the source (ionization) frequency. The terms on the right-hand side represent the forces due to the electric field, pressure gradient, and collisions respectively. The largest of these terms is the pressure gradient term.

\begin{figure}
    \includegraphics{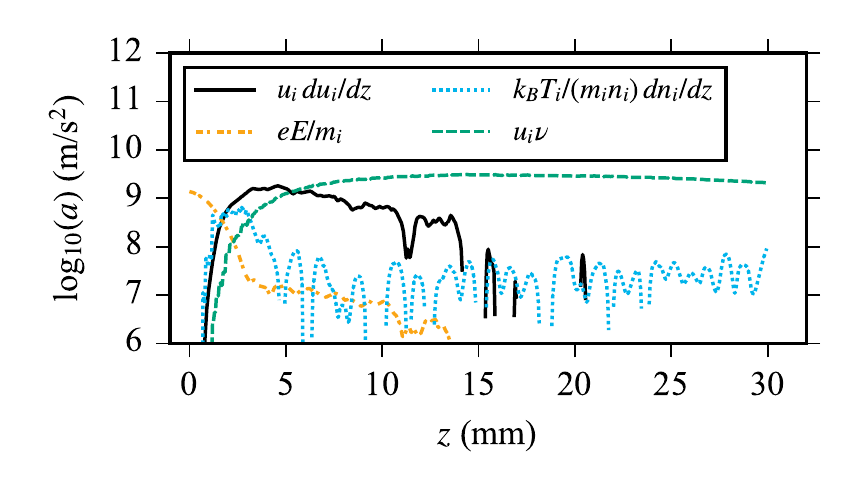}
    \caption{Calculation of the ion fluid terms along the axis of symmetry. Two-dimensional effects are excluded. The inertial term (RHS Eq.~\ref{eqn:com}) is the solid line, the electric field term (first term on LHS) is the dash-dotted line, the pressure term (second term on LHS) is the dotted line, and collisions (third term on LHS) are shown by the dashed line.}
    \label{fig:ion_fluid_terms}
\end{figure}

The ions are estimated to vary according to the Boltzmann density relation,
\begin{equation}
    \frac{eE}{k_BT_i} = \frac{1}{n}\frac{dn}{dz}.
    \label{eqn:boltz}
\end{equation}
This assumption depends on a number of factors which may not necessarily be met for all electron sheaths. Multi-dimensional effects, flow, and collisions may all reduce the applicability of this relation. Indeed, in the simulations conducted, this relation is not strictly applicable as the ions experience substantial collisional and inertial forces as seen in Figure~\ref{fig:ion_fluid_terms}.
The figure depicts the inertial (solid), pressure (dotted), electric field (dash-dotted), and collisional (dotted) fluid terms for the ions along the axis of symmetry, neglecting perpendicular components. These cases require significantly more complex analysis; an example of the treatment of the multi-dimensional case can be found in~\cite{Scheiner2015}. Therefore, this relation is assumed here strictly as a means of developing an initial estimate of electron sheath and presheath properties in the ideal case of minimal ion-neutral collisions and negligible flow. We note that no such assumption is made in the simulations.

\begin{figure}
    \includegraphics{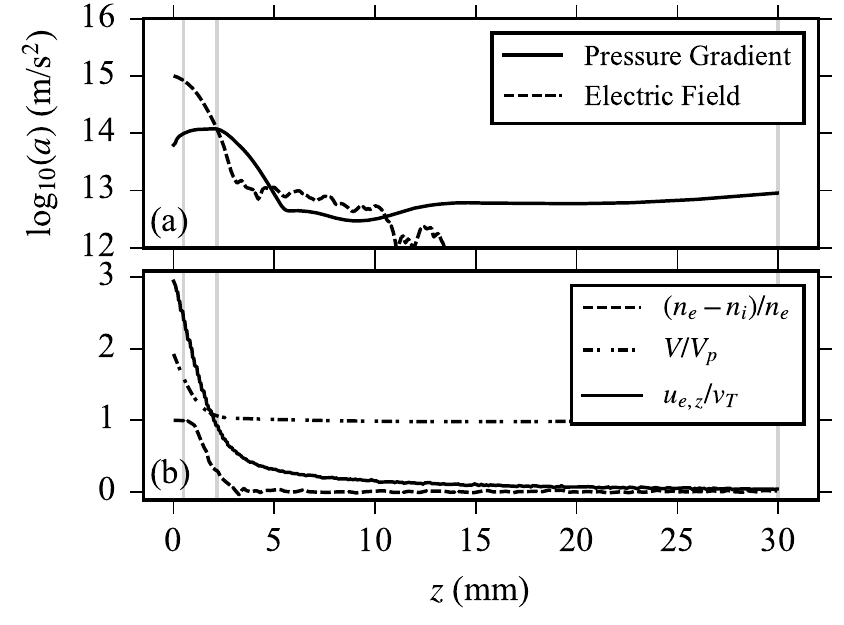}
    \caption{(a) Pressure gradient (solid line) and the electric field (dotted line) accelerations obtained from simulations. The solid gray lines indicate the locations where the \textsc{evdf}s in Fig.~\ref{fig:evdfs} were obtained, with the middle line indicating the location of the sheath edge. (b) Normalized charge density (dashed line), the electron fluid velocity (solid line), and potential (dash-dotted line) in the direction normal to the electrode surface, with respect to the distance from the electrode along $x=0.0$.}
    \label{fig:force_terms}
\end{figure}

Substituting this into Eq.~(\ref{eqn:com}) shows that the pressure gradient term is $T_e/T_i$ larger than the electric field. This is consistent with previous emissive probe measurements in a discharge of lower density ($10^8$ cm$^{-3}$ compared to $10^9$ cm$^{-3}$) which showed essentially no field past 4 mm from the electrode~\cite{Barnat2014}. Although the potential gradient in the presheath is small (characterized by $T_i$), the resulting pressure gradient drives a strong electron flow due to the density gradient that results.  This contrasts with the situation found in ion presheaths, where the electric field term exceeds the pressure gradient term by $T_e/T_i$, and the presheath potential drop is of the order of $T_e$ (rather than $T_i$). In this case, ions are accelerated ballistically by the electric field to a speed exceeding the sound speed at the sheath edge. The importance of the pressure gradient term with respect to the electric field is confirmed in the simulations. Figure~\ref{fig:force_terms}(a)
plots the pressure gradient (solid line) and electric field (dashed line) acceleration terms from Eq.~(\ref{eqn:com}) on a logarithmic scale, as a function of distance from the electrode.

% ...

The simulations indicate that the pressure gradient is the dominant acceleration mechanism from 2--5 mm and $z>11$ mm. The drop in pressure gradient between 5 mm and 11 mm coincides with a plateau in the simulated electron density near the sheath edge (similar to that seen in Fig.~\ref{fig:on_axis_densities}) and a plateau in electron temperature (not shown). Several factors may contribute to this plateau including a stagnation of ions as they approach the sheath potential barrier or a change in the convergence of the electron fluid. Past 11 mm, the pressure gradient continues to dominate the acceleration of the electron fluid up to the sheath edge.


The degree to which the electron fluid is accelerated can be calculated via the electron continuity equation
\begin{equation}
    \frac{d}{dz}(n_e u_e) = \nu_sn_e,
    \label{eqn:cont}
\end{equation}
and a common sheath criterion, which identifies the sheath edge as the location where $\sum q\,{}dn/dz \le 0$ \cite{Riemann1991}, where $q$ is the charge of the species. Dropping the collision terms, Eqs.~(\ref{eqn:com}) and (\ref{eqn:cont}) yield $d n_e/dz = e n_e E / (m_e u_e^2 - k_B T_e)$. From Eq.~(\ref{eqn:boltz}), and assuming quasineutrality, the sheath criterion provides an electron-sheath analog of the Bohm criterion,
\begin{equation}
    u_e \ge \sqrt{\frac{k_B \left(T_e + T_i\right) }{m_e}} \equiv v_T.
    \label{eqn:bohm}
\end{equation}
A similar derivation of the electron fluid speed at the sheath edge was previously obtained by Loizu, Dominski, Ricci, and Theiler~\cite{Loizu2012} This criterion demands a region of electron acceleration outside of the sheath region. 


Figure~\ref{fig:force_terms}(b), shows the charge density (dashed line), the electron fluid velocity (solid line), and the local potential (dash-dotted line). The electron fluid is found to reach a velocity of $0.92v_T$ by the sheath edge (as previously defined), in fair agreement with Eq.~(\ref{eqn:bohm}). A number of factors may contribute to the remaining discrepancy including the ambiguity in the definition of a precise sheath edge location \cite{Baalrud2015}, the use of a planar one-dimensional theory in describing a converging flow, the presence of adjacent sheaths, and the substantial fluctuations observed in the simulations near the sheath edge.

% ...

The fast electron flow creates a large differential streaming between electrons and ions that is expected to lead to electron-ion streaming instabilities \cite{Bellan2006} in the electron presheath and sheath. Indeed, Fig.~\ref{fig:flow_instability} shows substantial fluctuations of the electron sheath edge, but not in the case of the ion sheath simulation. The frequency of these fluctuations is observed to peak around 0.8 MHz comparable to the most unstable mode of ion-acoustic instability (ion plasma frequency) which would be 1.4 MHz for $n_i = 1.7\times 10^8$ cm$^{-3}$ ($z = 2.0$ mm). A more in-depth investigation of the ion density fluctuations shows good agrement with the ion-acoustic dispersion relation~\cite{Scheiner2015}. This observation is consistent with previous observations of ion acoustic waves excited by positive probes~\cite{Glanz1981} and current fluctuations in measurements using a segmented electrode~\cite{Barnat2014}. It should also be noted that the low frequency of these fluctuations suggest that they are distinct from previous instabilities observed near electron-collecting interfaces~\cite{Stenzel1988, Stenzel2011a, Stenzel2011b}.

In addition, it is observed that the ion flow obtains a radial component near the sheath edge, which may contribute to ion-ion two-stream instabilities. The presence of these strong fluctuations blurs the sheath edge location, as shown in Fig.~\ref{fig:flow_instability}. In a steady-state sense, a transition region is formed between presheath and sheath as a result of the fluctuations. The previous estimate of the sheath edge (at $z=2.0$ mm) was based on where the time-average charge density is reduced by $30\%$ (if instead an estimate based on the Child-Langmuir sheath thickness is used, Eq.~(13) of \cite{Hershkowitz2005}, an estimate of $z=$ 1.5 mm is obtained \footnote{In his paper, Hershkowitz identifies Eq.~(17) as the one to be used for electron sheaths, however this expression assumes only random thermal electron flux enters the sheath. Given the observed flow shift, Eq.~(13) is considered the more appropriate choice.}).


The length scale of an ion presheath is typically determined by collisional processes, and is estimated as the ratio of the ion flow speed to the collision frequency $l_i \simeq c_s/\nu$ for a specific process such as ionization.  The electron presheath length would be estimated to be $l_e \simeq v_T/\nu$. This implies that, for the same collision process, the electron presheath is much longer than the ion presheath, by a factor of
\begin{equation}
    \frac{l_e}{l_i} = \frac{v_T}{c_s}
                    = \sqrt{\frac{T_e + T_i}{m_e}\frac{m_i}{T_e}} \approx \sqrt{\frac{m_i}{m_e}}.
    \label{eqn:extent}
\end{equation}
As an example, $l_e/l_i \approx 270$ in argon or $l_e/l_i \approx 85$ in helium. However, this ratio is only applicable in the case of the same collision process. In the present case, it is believed that the ion and electron presheath lengths are likely governed by different processes, namely ion-neutral and electron-neutral collisions. This suggests $l_e/l_i \approx 3$ based on calculations of the mean free paths. We also note that in typical low temperature plasma experiments, these lengths may be constrained by the dimensions of the plasma rather than collisions. Figures~\ref{fig:on_axis_densities} and \ref{fig:flow_instability} support the suggestion that electron presheaths are much longer than ion presheaths. This is a very different picture than the assumption of a local phenomenon that is found in Langmuir probe theory.

\section{Conclusions}
    These novel properties of the electron sheath are surprising both because of how they differ from ion sheaths and because of their influence on the bulk plasma. The perturbations in electron density and flow caused by what would otherwise be considered a small electrode suggests that conventional models of electron sheaths need to be revisited. These fundamental physics results may also lead to useful new applications.

\begin{acknowledgments}
    This work was supported by the Office of Fusion Energy Science at the U.S.
Department of Energy under contracts DE-AC04-94SL85000 and DE-SC0001939. One of
the authors, BS, is supported by the U.S.~Department of Energy, Office of
Science, Office of Workforce Development for Teachers and Scients, Office of
Science Graduate Student Research Program (\textsc{scgsr}). The \textsc{scgsr}
program is administered by the Oak Ridge Institute for Science and Education for
the the DOE under Contract No.~DE-AC05-06OR23100.

\end{acknowledgments}

\bibliographystyle{apsrev4-1}
\bibliography{main}

\begin{thebibliography}{38}%
\makeatletter
\providecommand \@ifxundefined [1]{%
 \@ifx{#1\undefined}
}%
\providecommand \@ifnum [1]{%
 \ifnum #1\expandafter \@firstoftwo
 \else \expandafter \@secondoftwo
 \fi
}%
\providecommand \@ifx [1]{%
 \ifx #1\expandafter \@firstoftwo
 \else \expandafter \@secondoftwo
 \fi
}%
\providecommand \natexlab [1]{#1}%
\providecommand \enquote  [1]{``#1''}%
\providecommand \bibnamefont  [1]{#1}%
\providecommand \bibfnamefont [1]{#1}%
\providecommand \citenamefont [1]{#1}%
\providecommand \href@noop [0]{\@secondoftwo}%
\providecommand \href [0]{\begingroup \@sanitize@url \@href}%
\providecommand \@href[1]{\@@startlink{#1}\@@href}%
\providecommand \@@href[1]{\endgroup#1\@@endlink}%
\providecommand \@sanitize@url [0]{\catcode `\\12\catcode `\$12\catcode
  `\&12\catcode `\#12\catcode `\^12\catcode `\_12\catcode `\%12\relax}%
\providecommand \@@startlink[1]{}%
\providecommand \@@endlink[0]{}%
\providecommand \url  [0]{\begingroup\@sanitize@url \@url }%
\providecommand \@url [1]{\endgroup\@href {#1}{\urlprefix }}%
\providecommand \urlprefix  [0]{URL }%
\providecommand \Eprint [0]{\href }%
\providecommand \doibase [0]{http://dx.doi.org/}%
\providecommand \selectlanguage [0]{\@gobble}%
\providecommand \bibinfo  [0]{\@secondoftwo}%
\providecommand \bibfield  [0]{\@secondoftwo}%
\providecommand \translation [1]{[#1]}%
\providecommand \BibitemOpen [0]{}%
\providecommand \bibitemStop [0]{}%
\providecommand \bibitemNoStop [0]{.\EOS\space}%
\providecommand \EOS [0]{\spacefactor3000\relax}%
\providecommand \BibitemShut  [1]{\csname bibitem#1\endcsname}%
\let\auto@bib@innerbib\@empty
\bibitem [{\citenamefont {Langmuir}(1929)}]{Langmuir1929}%
  \BibitemOpen
  \bibfield  {author} {\bibinfo {author} {\bibfnamefont {I.}~\bibnamefont
  {Langmuir}},\ }\href {\doibase 10.1103/PhysRev.33.954} {\bibfield  {journal}
  {\bibinfo  {journal} {Phys. Rev.}\ }\textbf {\bibinfo {volume} {33}},\
  \bibinfo {pages} {954} (\bibinfo {year} {1929})}\BibitemShut {NoStop}%
\bibitem [{\citenamefont {Bacal}(2006)}]{Bacal2006}%
  \BibitemOpen
  \bibfield  {author} {\bibinfo {author} {\bibfnamefont {M.}~\bibnamefont
  {Bacal}},\ }\href {\doibase 10.1088/0029-5515/46/6/S05} {\bibfield  {journal}
  {\bibinfo  {journal} {Nucl. Fusion}\ }\textbf {\bibinfo {volume} {46}},\
  \bibinfo {pages} {S250} (\bibinfo {year} {2006})}\BibitemShut {NoStop}%
\bibitem [{\citenamefont {Longmier}\ \emph {et~al.}(2006)\citenamefont
  {Longmier}, \citenamefont {Baalrud},\ and\ \citenamefont
  {Hershkowitz}}]{Longmier2006}%
  \BibitemOpen
  \bibfield  {author} {\bibinfo {author} {\bibfnamefont {B.}~\bibnamefont
  {Longmier}}, \bibinfo {author} {\bibfnamefont {S.~D.}\ \bibnamefont
  {Baalrud}}, \ and\ \bibinfo {author} {\bibfnamefont {N.}~\bibnamefont
  {Hershkowitz}},\ }\href {\doibase 10.1063/1.2393164} {\bibfield  {journal}
  {\bibinfo  {journal} {Rev Sci. Instrum.}\ }\textbf {\bibinfo {volume} {77}}
  (\bibinfo {year} {2006}),\ 10.1063/1.2393164}\BibitemShut {NoStop}%
\bibitem [{\citenamefont {Theiler}\ \emph {et~al.}(2012)\citenamefont
  {Theiler}, \citenamefont {Furno}, \citenamefont {Loizu},\ and\ \citenamefont
  {Fasoli}}]{Theiler2012}%
  \BibitemOpen
  \bibfield  {author} {\bibinfo {author} {\bibfnamefont {C.}~\bibnamefont
  {Theiler}}, \bibinfo {author} {\bibfnamefont {I.}~\bibnamefont {Furno}},
  \bibinfo {author} {\bibfnamefont {J.}~\bibnamefont {Loizu}}, \ and\ \bibinfo
  {author} {\bibfnamefont {A.}~\bibnamefont {Fasoli}},\ }\href {\doibase
  10.1103/PhysRevLett.108.065005} {\bibfield  {journal} {\bibinfo  {journal}
  {Phys. Rev. Lett.}\ }\textbf {\bibinfo {volume} {108}},\ \bibinfo {pages}
  {065005} (\bibinfo {year} {2012})}\BibitemShut {NoStop}%
\bibitem [{\citenamefont {Law}\ \emph {et~al.}(1998)\citenamefont {Law},
  \citenamefont {Steel}, \citenamefont {Annaratone},\ and\ \citenamefont
  {Allen}}]{Law1998}%
  \BibitemOpen
  \bibfield  {author} {\bibinfo {author} {\bibfnamefont {D.~A.}\ \bibnamefont
  {Law}}, \bibinfo {author} {\bibfnamefont {W.~H.}\ \bibnamefont {Steel}},
  \bibinfo {author} {\bibfnamefont {B.~M.}\ \bibnamefont {Annaratone}}, \ and\
  \bibinfo {author} {\bibfnamefont {J.~E.}\ \bibnamefont {Allen}},\ }\href
  {\doibase 10.1103/PhysRevLett.80.4189} {\bibfield  {journal} {\bibinfo
  {journal} {Phys. Rev. Lett.}\ }\textbf {\bibinfo {volume} {80}},\ \bibinfo
  {pages} {4189} (\bibinfo {year} {1998})}\BibitemShut {NoStop}%
\bibitem [{\citenamefont {Richards}\ \emph {et~al.}(1994)\citenamefont
  {Richards}, \citenamefont {Uckan}, \citenamefont {Wootton}, \citenamefont
  {Carreras}, \citenamefont {Bengtson}, \citenamefont {Hurwitz}, \citenamefont
  {Li}, \citenamefont {Lin}, \citenamefont {Rowan}, \citenamefont {Tsui},
  \citenamefont {Sen},\ and\ \citenamefont {Uglum}}]{Richards1994}%
  \BibitemOpen
  \bibfield  {author} {\bibinfo {author} {\bibfnamefont {B.}~\bibnamefont
  {Richards}}, \bibinfo {author} {\bibfnamefont {T.}~\bibnamefont {Uckan}},
  \bibinfo {author} {\bibfnamefont {A.~J.}\ \bibnamefont {Wootton}}, \bibinfo
  {author} {\bibfnamefont {B.~A.}\ \bibnamefont {Carreras}}, \bibinfo {author}
  {\bibfnamefont {R.~D.}\ \bibnamefont {Bengtson}}, \bibinfo {author}
  {\bibfnamefont {P.}~\bibnamefont {Hurwitz}}, \bibinfo {author} {\bibfnamefont
  {G.~X.}\ \bibnamefont {Li}}, \bibinfo {author} {\bibfnamefont
  {H.}~\bibnamefont {Lin}}, \bibinfo {author} {\bibfnamefont {W.~L.}\
  \bibnamefont {Rowan}}, \bibinfo {author} {\bibfnamefont {H.~Y.~W.}\
  \bibnamefont {Tsui}}, \bibinfo {author} {\bibfnamefont {A.~K.}\ \bibnamefont
  {Sen}}, \ and\ \bibinfo {author} {\bibfnamefont {J.}~\bibnamefont {Uglum}},\
  }\href {\doibase 10.1063/1.870661} {\bibfield  {journal} {\bibinfo  {journal}
  {Phys. Plasmas}\ }\textbf {\bibinfo {volume} {1}},\ \bibinfo {pages} {1606}
  (\bibinfo {year} {1994})}\BibitemShut {NoStop}%
\bibitem [{\citenamefont {Campanell}(2013)}]{Campanell2013}%
  \BibitemOpen
  \bibfield  {author} {\bibinfo {author} {\bibfnamefont {M.~D.}\ \bibnamefont
  {Campanell}},\ }\href {\doibase 10.1103/PhysRevE.88.033103} {\bibfield
  {journal} {\bibinfo  {journal} {Phys. Rev. E}\ }\textbf {\bibinfo {volume}
  {88}},\ \bibinfo {pages} {033103} (\bibinfo {year} {2013})}\BibitemShut
  {NoStop}%
\bibitem [{\citenamefont {Wang}\ \emph {et~al.}(2006)\citenamefont {Wang},
  \citenamefont {Economou},\ and\ \citenamefont {Donnelly}}]{Wang2006}%
  \BibitemOpen
  \bibfield  {author} {\bibinfo {author} {\bibfnamefont {Q.}~\bibnamefont
  {Wang}}, \bibinfo {author} {\bibfnamefont {D.~J.}\ \bibnamefont {Economou}},
  \ and\ \bibinfo {author} {\bibfnamefont {V.~M.}\ \bibnamefont {Donnelly}},\
  }\href {\doibase 10.1063/1.2214591} {\bibfield  {journal} {\bibinfo
  {journal} {J. Appl. Phys.}\ }\textbf {\bibinfo {volume} {100}} (\bibinfo
  {year} {2006}),\ 10.1063/1.2214591}\BibitemShut {NoStop}%
\bibitem [{\citenamefont {Mahony}\ \emph {et~al.}(1997)\citenamefont {Mahony},
  \citenamefont {{Al Wazzan}},\ and\ \citenamefont {Graham}}]{Mahony1997}%
  \BibitemOpen
  \bibfield  {author} {\bibinfo {author} {\bibfnamefont {C.~M.~O.}\
  \bibnamefont {Mahony}}, \bibinfo {author} {\bibfnamefont {R.}~\bibnamefont
  {{Al Wazzan}}}, \ and\ \bibinfo {author} {\bibfnamefont {W.~G.}\ \bibnamefont
  {Graham}},\ }\href {\doibase 10.1063/1.119808} {\bibfield  {journal}
  {\bibinfo  {journal} {Appl. Phys. Lett.}\ }\textbf {\bibinfo {volume} {71}},\
  \bibinfo {pages} {608} (\bibinfo {year} {1997})}\BibitemShut {NoStop}%
\bibitem [{\citenamefont {Sanmartin}\ \emph {et~al.}(1993)\citenamefont
  {Sanmartin}, \citenamefont {Mart\'{\i}nez-S\'{a}nchez},\ and\ \citenamefont
  {Ahedo}}]{Sanmartin1993}%
  \BibitemOpen
  \bibfield  {author} {\bibinfo {author} {\bibfnamefont {J.~R.}\ \bibnamefont
  {Sanmartin}}, \bibinfo {author} {\bibfnamefont {M.}~\bibnamefont
  {Mart\'{\i}nez-S\'{a}nchez}}, \ and\ \bibinfo {author} {\bibfnamefont
  {E.}~\bibnamefont {Ahedo}},\ }\href@noop {} {\bibfield  {journal} {\bibinfo
  {journal} {J. Propuls. Power}\ }\textbf {\bibinfo {volume} {9}},\ \bibinfo
  {pages} {353} (\bibinfo {year} {1993})}\BibitemShut {NoStop}%
\bibitem [{\citenamefont {Poppe}\ \emph {et~al.}(2011)\citenamefont {Poppe},
  \citenamefont {Halekas},\ and\ \citenamefont {Hor\"{a}nyi}}]{Poppe2011}%
  \BibitemOpen
  \bibfield  {author} {\bibinfo {author} {\bibfnamefont {A.}~\bibnamefont
  {Poppe}}, \bibinfo {author} {\bibfnamefont {J.~S.}\ \bibnamefont {Halekas}},
  \ and\ \bibinfo {author} {\bibfnamefont {M.}~\bibnamefont {Hor\"{a}nyi}},\
  }\href {\doibase 10.1029/2010GL046119} {\bibfield  {journal} {\bibinfo
  {journal} {Geophy. Res. Lett.}\ }\textbf {\bibinfo {volume} {38}},\ \bibinfo
  {pages} {L02103} (\bibinfo {year} {2011})}\BibitemShut {NoStop}%
\bibitem [{\citenamefont {Yip}\ and\ \citenamefont
  {Hershkowitz}(2015)}]{Yip2015}%
  \BibitemOpen
  \bibfield  {author} {\bibinfo {author} {\bibfnamefont {C.-S.}\ \bibnamefont
  {Yip}}\ and\ \bibinfo {author} {\bibfnamefont {N.}~\bibnamefont
  {Hershkowitz}},\ }\href {\doibase 10.1088/0963-0252/24/3/034004} {\bibfield
  {journal} {\bibinfo  {journal} {Plasma Sources Sci. Technol.}\ }\textbf
  {\bibinfo {volume} {24}},\ \bibinfo {pages} {034004} (\bibinfo {year}
  {2015})}\BibitemShut {NoStop}%
\bibitem [{\citenamefont {Zweben}\ \emph {et~al.}(2009)\citenamefont {Zweben},
  \citenamefont {Maqueda}, \citenamefont {Roquemore}, \citenamefont {Bush},
  \citenamefont {Kaita}, \citenamefont {Marsala}, \citenamefont {Raitses},
  \citenamefont {Cohen},\ and\ \citenamefont {Ryutov}}]{Zweben2009}%
  \BibitemOpen
  \bibfield  {author} {\bibinfo {author} {\bibfnamefont {S.~J.}\ \bibnamefont
  {Zweben}}, \bibinfo {author} {\bibfnamefont {R.~J.}\ \bibnamefont {Maqueda}},
  \bibinfo {author} {\bibfnamefont {A.~L.}\ \bibnamefont {Roquemore}}, \bibinfo
  {author} {\bibfnamefont {C.~E.}\ \bibnamefont {Bush}}, \bibinfo {author}
  {\bibfnamefont {R.}~\bibnamefont {Kaita}}, \bibinfo {author} {\bibfnamefont
  {R.~J.}\ \bibnamefont {Marsala}}, \bibinfo {author} {\bibfnamefont
  {Y.}~\bibnamefont {Raitses}}, \bibinfo {author} {\bibfnamefont {R.~H.}\
  \bibnamefont {Cohen}}, \ and\ \bibinfo {author} {\bibfnamefont {D.~D.}\
  \bibnamefont {Ryutov}},\ }\href {\doibase 10.1088/0741-3335/51/10/105012}
  {\bibfield  {journal} {\bibinfo  {journal} {Plasma Phys. Control. Fusion}\
  }\textbf {\bibinfo {volume} {51}},\ \bibinfo {pages} {105012} (\bibinfo
  {year} {2009})}\BibitemShut {NoStop}%
\bibitem [{\citenamefont {Bohm}(1949)}]{Bohm1949}%
  \BibitemOpen
  \bibfield  {author} {\bibinfo {author} {\bibfnamefont {D.}~\bibnamefont
  {Bohm}},\ }in\ \href@noop {} {\emph {\bibinfo {booktitle} {The
  Characteristics of Electrical Discharges in Magnetic Fields}}},\ \bibinfo
  {editor} {edited by\ \bibinfo {editor} {\bibfnamefont {A.}~\bibnamefont
  {Guthrie}}\ and\ \bibinfo {editor} {\bibfnamefont {R.~K.}\ \bibnamefont
  {Wakerling}}}\ (\bibinfo  {publisher} {McGraw-Hill},\ \bibinfo {address} {New
  York, NY},\ \bibinfo {year} {1949})\ Chap.~\bibinfo {chapter} {3},
  p.~\bibinfo {pages} {77}\BibitemShut {NoStop}%
\bibitem [{\citenamefont {Baalrud}\ \emph {et~al.}(2007)\citenamefont
  {Baalrud}, \citenamefont {Hershkowitz},\ and\ \citenamefont
  {Longmier}}]{Baalrud2007}%
  \BibitemOpen
  \bibfield  {author} {\bibinfo {author} {\bibfnamefont {S.~D.}\ \bibnamefont
  {Baalrud}}, \bibinfo {author} {\bibfnamefont {N.}~\bibnamefont
  {Hershkowitz}}, \ and\ \bibinfo {author} {\bibfnamefont {B.}~\bibnamefont
  {Longmier}},\ }\href {\doibase 10.1063/1.2722262} {\bibfield  {journal}
  {\bibinfo  {journal} {Phys. Plasmas}\ }\textbf {\bibinfo {volume} {14}},\
  \bibinfo {pages} {042109} (\bibinfo {year} {2007})}\BibitemShut {NoStop}%
\bibitem [{\citenamefont {Barnat}\ and\ \citenamefont
  {Frederickson}(2010)}]{Barnat2010}%
  \BibitemOpen
  \bibfield  {author} {\bibinfo {author} {\bibfnamefont {E.~V.}\ \bibnamefont
  {Barnat}}\ and\ \bibinfo {author} {\bibfnamefont {K.}~\bibnamefont
  {Frederickson}},\ }\href {\doibase 10.1088/0963-0252/19/5/055015} {\bibfield
  {journal} {\bibinfo  {journal} {Plasma Sources Sci. T.}\ }\textbf {\bibinfo
  {volume} {19}},\ \bibinfo {pages} {055015} (\bibinfo {year}
  {2010})}\BibitemShut {NoStop}%
\bibitem [{\citenamefont {Weatherford}\ \emph {et~al.}(2012)\citenamefont
  {Weatherford}, \citenamefont {Barnat},\ and\ \citenamefont
  {Foster}}]{Weatherford2012b}%
  \BibitemOpen
  \bibfield  {author} {\bibinfo {author} {\bibfnamefont {B.~R.}\ \bibnamefont
  {Weatherford}}, \bibinfo {author} {\bibfnamefont {E.~V.}\ \bibnamefont
  {Barnat}}, \ and\ \bibinfo {author} {\bibfnamefont {J.~E.}\ \bibnamefont
  {Foster}},\ }\href {\doibase 10.1088/0963-0252/21/5/055030} {\bibfield
  {journal} {\bibinfo  {journal} {Plasma Sources Sci. Technol.}\ }\textbf
  {\bibinfo {volume} {21}},\ \bibinfo {pages} {055030} (\bibinfo {year}
  {2012})}\BibitemShut {NoStop}%
\bibitem [{\citenamefont {Riemann}(1991)}]{Riemann1991}%
  \BibitemOpen
  \bibfield  {author} {\bibinfo {author} {\bibfnamefont {K.-U.}\ \bibnamefont
  {Riemann}},\ }\href {\doibase 10.1088/0022-3727/24/4/001} {\bibfield
  {journal} {\bibinfo  {journal} {J. Phys. D}\ }\textbf {\bibinfo {volume}
  {24}},\ \bibinfo {pages} {493} (\bibinfo {year} {1991})}\BibitemShut
  {NoStop}%
\bibitem [{\citenamefont {Phelps}(2002)}]{Phelps2002}%
  \BibitemOpen
  \bibfield  {author} {\bibinfo {author} {\bibfnamefont {A.~V.}\ \bibnamefont
  {Phelps}},\ }\href {http://www.lxcat.net} {\enquote {\bibinfo {title} {{He+
  in He Cross Sections}},}\ } (\bibinfo {year} {2002})\BibitemShut {NoStop}%
\bibitem [{\citenamefont {Heroux}\ \emph {et~al.}(2005)\citenamefont {Heroux},
  \citenamefont {Phipps}, \citenamefont {Salinger}, \citenamefont {Thornquist},
  \citenamefont {Tuminaro}, \citenamefont {Willenbring}, \citenamefont
  {Williams}, \citenamefont {Stanley}, \citenamefont {Bartlett}, \citenamefont
  {Howle}, \citenamefont {Hoekstra}, \citenamefont {Hu}, \citenamefont {Kolda},
  \citenamefont {Lehoucq}, \citenamefont {Long},\ and\ \citenamefont
  {Pawlowski}}]{Heroux2005}%
  \BibitemOpen
  \bibfield  {author} {\bibinfo {author} {\bibfnamefont {M.~A.}\ \bibnamefont
  {Heroux}}, \bibinfo {author} {\bibfnamefont {E.~T.}\ \bibnamefont {Phipps}},
  \bibinfo {author} {\bibfnamefont {A.~G.}\ \bibnamefont {Salinger}}, \bibinfo
  {author} {\bibfnamefont {H.~K.}\ \bibnamefont {Thornquist}}, \bibinfo
  {author} {\bibfnamefont {R.~S.}\ \bibnamefont {Tuminaro}}, \bibinfo {author}
  {\bibfnamefont {J.~M.}\ \bibnamefont {Willenbring}}, \bibinfo {author}
  {\bibfnamefont {A.}~\bibnamefont {Williams}}, \bibinfo {author}
  {\bibfnamefont {K.~S.}\ \bibnamefont {Stanley}}, \bibinfo {author}
  {\bibfnamefont {R.~A.}\ \bibnamefont {Bartlett}}, \bibinfo {author}
  {\bibfnamefont {V.~E.}\ \bibnamefont {Howle}}, \bibinfo {author}
  {\bibfnamefont {R.~J.}\ \bibnamefont {Hoekstra}}, \bibinfo {author}
  {\bibfnamefont {J.~J.}\ \bibnamefont {Hu}}, \bibinfo {author} {\bibfnamefont
  {T.~G.}\ \bibnamefont {Kolda}}, \bibinfo {author} {\bibfnamefont {R.~B.}\
  \bibnamefont {Lehoucq}}, \bibinfo {author} {\bibfnamefont {K.~R.}\
  \bibnamefont {Long}}, \ and\ \bibinfo {author} {\bibfnamefont {R.~P.}\
  \bibnamefont {Pawlowski}},\ }\href {\doibase 10.1145/1089014.1089021}
  {\bibfield  {journal} {\bibinfo  {journal} {ACM T. Math. Software}\ }\textbf
  {\bibinfo {volume} {31}},\ \bibinfo {pages} {397} (\bibinfo {year}
  {2005})}\BibitemShut {NoStop}%
\bibitem [{\citenamefont {Spreiter}\ and\ \citenamefont
  {Walter}(1999)}]{Spreiter1999}%
  \BibitemOpen
  \bibfield  {author} {\bibinfo {author} {\bibfnamefont {Q.}~\bibnamefont
  {Spreiter}}\ and\ \bibinfo {author} {\bibfnamefont {M.}~\bibnamefont
  {Walter}},\ }\href {\doibase 10.1006/jcph.1999.6237} {\bibfield  {journal}
  {\bibinfo  {journal} {J. Comp. Phys.}\ }\textbf {\bibinfo {volume} {152}},\
  \bibinfo {pages} {102} (\bibinfo {year} {1999})}\BibitemShut {NoStop}%
\bibitem [{\citenamefont {Timko}\ \emph {et~al.}(2012)\citenamefont {Timko},
  \citenamefont {Crozier}, \citenamefont {Hopkins}, \citenamefont {Matyash},\
  and\ \citenamefont {Schneider}}]{Timko2012}%
  \BibitemOpen
  \bibfield  {author} {\bibinfo {author} {\bibfnamefont {H.}~\bibnamefont
  {Timko}}, \bibinfo {author} {\bibfnamefont {P.~S.}\ \bibnamefont {Crozier}},
  \bibinfo {author} {\bibfnamefont {M.~M.}\ \bibnamefont {Hopkins}}, \bibinfo
  {author} {\bibfnamefont {K.}~\bibnamefont {Matyash}}, \ and\ \bibinfo
  {author} {\bibfnamefont {R.}~\bibnamefont {Schneider}},\ }\href {\doibase
  10.1002/ctpp.201100051} {\bibfield  {journal} {\bibinfo  {journal} {Rev. Sci.
  Instrum.}\ }\textbf {\bibinfo {volume} {65}},\ \bibinfo {pages} {140}
  (\bibinfo {year} {2012})}\BibitemShut {NoStop}%
\bibitem [{\citenamefont {Hopkins}\ \emph {et~al.}(2016)\citenamefont
  {Hopkins}, \citenamefont {Yee}, \citenamefont {Baalrud},\ and\ \citenamefont
  {Barnat}}]{Hopkins2016}%
  \BibitemOpen
  \bibfield  {author} {\bibinfo {author} {\bibfnamefont {M.~M.}\ \bibnamefont
  {Hopkins}}, \bibinfo {author} {\bibfnamefont {B.~T.}\ \bibnamefont {Yee}},
  \bibinfo {author} {\bibfnamefont {S.~D.}\ \bibnamefont {Baalrud}}, \ and\
  \bibinfo {author} {\bibfnamefont {E.~V.}\ \bibnamefont {Barnat}},\ }\href
  {\doibase 10.1063/1.4953896} {\bibfield  {journal} {\bibinfo  {journal}
  {Phys. Plasmas}\ }\textbf {\bibinfo {volume} {23}},\ \bibinfo {pages}
  {063519} (\bibinfo {year} {2016})}\BibitemShut {NoStop}%
\bibitem [{\citenamefont {Baalrud}\ \emph {et~al.}(2015)\citenamefont
  {Baalrud}, \citenamefont {Scheiner}, \citenamefont {Yee}, \citenamefont
  {Hopkins},\ and\ \citenamefont {Barnat}}]{Baalrud2015}%
  \BibitemOpen
  \bibfield  {author} {\bibinfo {author} {\bibfnamefont {S.~D.}\ \bibnamefont
  {Baalrud}}, \bibinfo {author} {\bibfnamefont {B.}~\bibnamefont {Scheiner}},
  \bibinfo {author} {\bibfnamefont {B.~T.}\ \bibnamefont {Yee}}, \bibinfo
  {author} {\bibfnamefont {M.~M.}\ \bibnamefont {Hopkins}}, \ and\ \bibinfo
  {author} {\bibfnamefont {E.~V.}\ \bibnamefont {Barnat}},\ }\href {\doibase
  10.1088/0741-3335/57/4/044003} {\bibfield  {journal} {\bibinfo  {journal}
  {Plasma Phys. Contr. F.}\ }\textbf {\bibinfo {volume} {57}},\ \bibinfo
  {pages} {044003} (\bibinfo {year} {2015})}\BibitemShut {NoStop}%
\bibitem [{\citenamefont {Barnat}\ \emph {et~al.}(2014)\citenamefont {Barnat},
  \citenamefont {Laity},\ and\ \citenamefont {Baalrud}}]{Barnat2014}%
  \BibitemOpen
  \bibfield  {author} {\bibinfo {author} {\bibfnamefont {E.~V.}\ \bibnamefont
  {Barnat}}, \bibinfo {author} {\bibfnamefont {G.~R.}\ \bibnamefont {Laity}}, \
  and\ \bibinfo {author} {\bibfnamefont {S.~D.}\ \bibnamefont {Baalrud}},\
  }\href {\doibase 10.1063/1.4897927} {\bibfield  {journal} {\bibinfo
  {journal} {Phys. Plasmas}\ }\textbf {\bibinfo {volume} {21}},\ \bibinfo
  {pages} {103512} (\bibinfo {year} {2014})}\BibitemShut {NoStop}%
\bibitem [{\citenamefont {Scheiner}\ \emph {et~al.}(2015)\citenamefont
  {Scheiner}, \citenamefont {Baalrud}, \citenamefont {Yee}, \citenamefont
  {Hopkins},\ and\ \citenamefont {Barnat}}]{Scheiner2015}%
  \BibitemOpen
  \bibfield  {author} {\bibinfo {author} {\bibfnamefont {B.}~\bibnamefont
  {Scheiner}}, \bibinfo {author} {\bibfnamefont {S.~D.}\ \bibnamefont
  {Baalrud}}, \bibinfo {author} {\bibfnamefont {B.~T.}\ \bibnamefont {Yee}},
  \bibinfo {author} {\bibfnamefont {M.~M.}\ \bibnamefont {Hopkins}}, \ and\
  \bibinfo {author} {\bibfnamefont {E.~V.}\ \bibnamefont {Barnat}},\ }\href
  {\doibase 10.1063/1.4939024} {\bibfield  {journal} {\bibinfo  {journal}
  {Phys. Plasmas}\ }\textbf {\bibinfo {volume} {22}} (\bibinfo {year} {2015}),\
  10.1063/1.4939024}\BibitemShut {NoStop}%
\bibitem [{\citenamefont {Mott-Smith}\ and\ \citenamefont
  {Langmuir}(1926)}]{Mott-Smith1926}%
  \BibitemOpen
  \bibfield  {author} {\bibinfo {author} {\bibfnamefont {H.}~\bibnamefont
  {Mott-Smith}}\ and\ \bibinfo {author} {\bibfnamefont {I.}~\bibnamefont
  {Langmuir}},\ }\href {\doibase 10.1103/PhysRev.28.727} {\bibfield  {journal}
  {\bibinfo  {journal} {Phys. Rev.}\ }\textbf {\bibinfo {volume} {28}},\
  \bibinfo {pages} {727} (\bibinfo {year} {1926})}\BibitemShut {NoStop}%
\bibitem [{\citenamefont {Medicus}(1961)}]{Medicus1961}%
  \BibitemOpen
  \bibfield  {author} {\bibinfo {author} {\bibfnamefont {G.}~\bibnamefont
  {Medicus}},\ }\href {\doibase 10.1063/1.1728342} {\bibfield  {journal}
  {\bibinfo  {journal} {J. Appl. Phys.}\ }\textbf {\bibinfo {volume} {32}},\
  \bibinfo {pages} {2512} (\bibinfo {year} {1961})}\BibitemShut {NoStop}%
\bibitem [{\citenamefont {Hershkowitz}(2005)}]{Hershkowitz2005}%
  \BibitemOpen
  \bibfield  {author} {\bibinfo {author} {\bibfnamefont {N.}~\bibnamefont
  {Hershkowitz}},\ }\href {\doibase 10.1063/1.1887189} {\bibfield  {journal}
  {\bibinfo  {journal} {Phys. Plasmas}\ }\textbf {\bibinfo {volume} {12}},\
  \bibinfo {pages} {055502} (\bibinfo {year} {2005})}\BibitemShut {NoStop}%
\bibitem [{\citenamefont {Montgomery}\ and\ \citenamefont
  {Nielsen}(1970)}]{Montgomery1970}%
  \BibitemOpen
  \bibfield  {author} {\bibinfo {author} {\bibfnamefont {D.}~\bibnamefont
  {Montgomery}}\ and\ \bibinfo {author} {\bibfnamefont {C.~W.}\ \bibnamefont
  {Nielsen}},\ }\href {\doibase 10.1063/1.1693081} {\bibfield  {journal}
  {\bibinfo  {journal} {Phys. Fluids}\ }\textbf {\bibinfo {volume} {13}},\
  \bibinfo {pages} {1405} (\bibinfo {year} {1970})}\BibitemShut {NoStop}%
\bibitem [{\citenamefont {Scheiner}\ \emph {et~al.}(2016)\citenamefont
  {Scheiner}, \citenamefont {Baalrud}, \citenamefont {Hopkins}, \citenamefont
  {Yee},\ and\ \citenamefont {Barnat}}]{Scheiner2016}%
  \BibitemOpen
  \bibfield  {author} {\bibinfo {author} {\bibfnamefont {B.}~\bibnamefont
  {Scheiner}}, \bibinfo {author} {\bibfnamefont {S.~D.}\ \bibnamefont
  {Baalrud}}, \bibinfo {author} {\bibfnamefont {M.~M.}\ \bibnamefont
  {Hopkins}}, \bibinfo {author} {\bibfnamefont {B.~T.}\ \bibnamefont {Yee}}, \
  and\ \bibinfo {author} {\bibfnamefont {E.~V.}\ \bibnamefont {Barnat}},\
  }\href {http://arxiv.org/abs/1604.08251} {\ ,\ \bibinfo {pages} {1} (\bibinfo
  {year} {2016})},\ \Eprint {http://arxiv.org/abs/1604.08251}
  {arXiv:1604.08251} \BibitemShut {NoStop}%
\bibitem [{\citenamefont {Loizu}\ \emph {et~al.}(2012)\citenamefont {Loizu},
  \citenamefont {Dominski}, \citenamefont {Ricci},\ and\ \citenamefont
  {Theiler}}]{Loizu2012}%
  \BibitemOpen
  \bibfield  {author} {\bibinfo {author} {\bibfnamefont {J.}~\bibnamefont
  {Loizu}}, \bibinfo {author} {\bibfnamefont {J.}~\bibnamefont {Dominski}},
  \bibinfo {author} {\bibfnamefont {P.}~\bibnamefont {Ricci}}, \ and\ \bibinfo
  {author} {\bibfnamefont {C.}~\bibnamefont {Theiler}},\ }\href {\doibase
  10.1063/1.4745863} {\bibfield  {journal} {\bibinfo  {journal} {Phys.
  Plasmas}\ }\textbf {\bibinfo {volume} {19}},\ \bibinfo {pages} {083507}
  (\bibinfo {year} {2012})}\BibitemShut {NoStop}%
\bibitem [{\citenamefont {Bellan}(2006)}]{Bellan2006}%
  \BibitemOpen
  \bibfield  {author} {\bibinfo {author} {\bibfnamefont {P.}~\bibnamefont
  {Bellan}},\ }in\ \href@noop {} {\emph {\bibinfo {booktitle} {Fundamentals of
  Plasma Physics}}}\ (\bibinfo  {publisher} {Cambridge University Press},\
  \bibinfo {address} {Cambridge, UK},\ \bibinfo {year} {2006})\ p.\ \bibinfo
  {pages} {177}\BibitemShut {NoStop}%
\bibitem [{\citenamefont {Glanz}\ and\ \citenamefont
  {Hershkowitz}(1981)}]{Glanz1981}%
  \BibitemOpen
  \bibfield  {author} {\bibinfo {author} {\bibfnamefont {J.}~\bibnamefont
  {Glanz}}\ and\ \bibinfo {author} {\bibfnamefont {N.}~\bibnamefont
  {Hershkowitz}},\ }\href {\doibase 10.1088/0032-1028/23/4/005} {\bibfield
  {journal} {\bibinfo  {journal} {Plasma Phys.}\ }\textbf {\bibinfo {volume}
  {23}},\ \bibinfo {pages} {325} (\bibinfo {year} {1981})}\BibitemShut
  {NoStop}%
\bibitem [{\citenamefont {Stenzel}(1988)}]{Stenzel1988}%
  \BibitemOpen
  \bibfield  {author} {\bibinfo {author} {\bibfnamefont {R.~L.}\ \bibnamefont
  {Stenzel}},\ }\href {\doibase 10.1103/PhysRevLett.60.704} {\bibfield
  {journal} {\bibinfo  {journal} {Phys. Rev. Lett.}\ }\textbf {\bibinfo
  {volume} {60}},\ \bibinfo {pages} {704} (\bibinfo {year} {1988})}\BibitemShut
  {NoStop}%
\bibitem [{\citenamefont {Stenzel}\ \emph
  {et~al.}(2011{\natexlab{a}})\citenamefont {Stenzel}, \citenamefont
  {Gruenwald}, \citenamefont {Ionita},\ and\ \citenamefont
  {Schrittwieser}}]{Stenzel2011a}%
  \BibitemOpen
  \bibfield  {author} {\bibinfo {author} {\bibfnamefont {R.~L.}\ \bibnamefont
  {Stenzel}}, \bibinfo {author} {\bibfnamefont {J.}~\bibnamefont {Gruenwald}},
  \bibinfo {author} {\bibfnamefont {C.}~\bibnamefont {Ionita}}, \ and\ \bibinfo
  {author} {\bibfnamefont {R.}~\bibnamefont {Schrittwieser}},\ }\href {\doibase
  10.1063/1.3601858} {\bibfield  {journal} {\bibinfo  {journal} {Phys.
  Plasmas}\ }\textbf {\bibinfo {volume} {18}},\ \bibinfo {pages} {062112}
  (\bibinfo {year} {2011}{\natexlab{a}})}\BibitemShut {NoStop}%
\bibitem [{\citenamefont {Stenzel}\ \emph
  {et~al.}(2011{\natexlab{b}})\citenamefont {Stenzel}, \citenamefont
  {Gruenwald}, \citenamefont {Ionita},\ and\ \citenamefont
  {Schrittwieser}}]{Stenzel2011b}%
  \BibitemOpen
  \bibfield  {author} {\bibinfo {author} {\bibfnamefont {R.~L.}\ \bibnamefont
  {Stenzel}}, \bibinfo {author} {\bibfnamefont {J.}~\bibnamefont {Gruenwald}},
  \bibinfo {author} {\bibfnamefont {C.}~\bibnamefont {Ionita}}, \ and\ \bibinfo
  {author} {\bibfnamefont {R.}~\bibnamefont {Schrittwieser}},\ }\href {\doibase
  10.1063/1.3601860} {\bibfield  {journal} {\bibinfo  {journal} {Phys.
  Plasmas}\ }\textbf {\bibinfo {volume} {18}},\ \bibinfo {pages} {062113}
  (\bibinfo {year} {2011}{\natexlab{b}})}\BibitemShut {NoStop}%
\bibitem [{Note1()}]{Note1}%
  \BibitemOpen
  \bibinfo {note} {In his paper, Hershkowitz identifies Eq.~(17) as the one to
  be used for electron sheaths, however this expression assumes only random
  thermal electron flux enters the sheath. Given the observed flow shift,
  Eq.~(13) is considered the more appropriate choice.}\BibitemShut {Stop}%
\end{thebibliography}%

\end{document}